\documentclass[twocolumn,aps,prl,showpacs]{revtex4}

\usepackage{graphicx}
\usepackage{amsmath,verbatim}

\begin{document}
\title{Enhancing the Violation of the Einstein-Podolsky-Rosen Local Realism by
Quantum Hyper-entanglement}
\author{Marco Barbieri$^1$, Francesco De Martini$^1$, Paolo Mataloni$^1$, Giuseppe
Vallone$^2$, Ad\'{a}n Cabello$^3$}
\address{$^1$Dipartimento di Fisica dell'Universit\`a ''La Sapienza'' and\\
Consorzio Nazionale Interuniversitario per le Scienze Fisiche della Materia, \\
00185 Roma, Italy\\
$^2$Dipartimento di Fisica Teorica Universit\`a di Torino and\\
Istituto Nazionale di Fisica Nucleare - sezione di Torino - 10125 Torino, Italy\\
$^3$Departamento de F\'{\i}sica Aplicada II, Universidad de Sevilla,
41012 Sevilla, Spain}


\begin{abstract}
Mermin's observation [Phys. Rev. Lett. {\bf 65}, 1838 (1990)] that
the magnitude of the violation of local realism, defined as the
ratio between the quantum prediction and the classical bound, can
grow exponentially with the size of the system is demonstrated using
two-photon hyper-entangled states entangled in polarization and path
degrees of freedom, and local measurements of polarization and path
simultaneously.
\end{abstract}



\maketitle


Einstein, Podolsky, and Rosen (EPR) \cite{1} believed that the
results of experiments on a system $A$ which can be remotely
predicted from the results of spacelike separated experiments on a
system $B$ were determined only by $A$'s local properties. Bell's
discovery that this is not the case \cite{2} meant a spectacular
departure from the local realistic view of the world.

Not so long ago, it was thought that the magnitude of the violation
of local realism, defined as the ratio between the quantum
prediction and the classical bound, decreases as the size (i.e.,
number $n$ of particles and/or the number $N$ of internal degrees of
freedom) grows, as a manifestation of some intrinsic aspect of the
transition from quantum to classical behavior \cite{3}. Later, it
was found that this magnitude can remain constant as $N$ increases \cite{4,5,6}.
This constant behavior occurs in most bipartite $N$-level Bell inequalities
\cite{7}. However, there is some evidence that bipartite three-level
systems can provide stronger violations of local realism than
bipartite two-level systems \cite{8}.

It was Mermin \cite{9} who showed that the ratio between the quantum
prediction and the classical bound can grow as $2^{(n-1)/2}$ in the case of
$n$-qubit systems prepared in Greenberger-Horne-Zeilinger (GHZ) states \cite{10}.
However, this effect is difficult to observe in real experiments because
it is difficult to produce GHZ states with $n>4$ \cite{11,12,13,14} and
because $n$-particle GHZ states suffer a decoherence that also grows
with $n$ \cite{15}.

The aim of this Letter is to describe an experiment in which Mermin's
growing-with-size quantum nonlocality effect is observed. The experiment is
based on two ingredients. On one hand, on a Bell inequality
derived but from the EPR
criterion for elements of reality \cite{1} applied to higher dimensional local
subsystems, where two compatible observables of the same particle can be regarded as
simultaneous EPR elements of reality if
the result of measuring each of them can be remotely predicted with
certainty and this prediction is independent of
which other compatible observables are
measured simultaneously. Examples of this type of Bell inequalities can be found in
\cite{16,17,18,19}. Some of them have been recently tested in real
experiments \cite{20,21}.
A detailed discussion of these Bell inequalities based on the EPR elements of reality can be found in \cite{22}.
The second key ingredient is {\it hyper-entanglement} (i.e., entanglement involving
different degrees of freedom \cite{23,24}), and particularly the possibility
of producing double \cite{20,21,25} (and, eventually, triple \cite{26})
Bell hyper-entangled states. Under some assumptions, hyper-entanglement allows us to replace $n$
two-level systems by two $N$-level systems, which significatively reduces
the decoherence problems, simplifies the task of achieving space-like
separation between measurements, and dramatically increases the efficiency
of detecting EPR elements of reality (since each photodetection reveals
$\log _{2}N$ elements of reality) \cite{22}. While previous tests of Bell
inequalities using hyper-entangled states have confirmed entanglement in
each of the degrees of freedom separately by using different local setups
for each degree of freedom \cite{25,26}, the local measurements in our
experiment are designed to show entanglement in all degrees of freedom
using the same setup.

Our experiment is based on the properties of the two-photon hyper-entangled
state exhibiting entanglement both in polarization and momentum {\bf k}
degrees of freedom,
\begin{equation}
\left|\Psi\right\rangle =\frac{1}{2}\left( |H\rangle _{u}|H\rangle
_{d}-|V\rangle _{u}|V\rangle _{d}\right) \otimes \left( |l\rangle
_{u}|r\rangle _{d}+|r\rangle _{u}|l\rangle _{d}\right) \text{,}
\label{hyper-ent}
\end{equation}
where $\left| H\right\rangle _{j}$ and $\left| V\right\rangle _{j}$
represent horizontal and vertical polarization, and $\left| l\right\rangle
_{j}$ and $\left| r\right\rangle _{j}$ denote two orthonormal path states,
i.e. {\bf k}-modes, for photon-$j$ ($j=u,d$). In the above expression, $l$ ($r$)
and $u$ ($d$) correspond to the {\it left} ({\it right}) and {\it up}
({\it down}) sides of the {\it entanglement ring} ({\it e-ring}) of the
parametric source (see Fig. 1a).

In the first
part of the experiment we demonstrate that the state (\ref{hyper-ent})
violates two Bell inequalities, for polarization and path, separately (a
similar experiment has been performed recently both by us and the Urbana
group \cite{25,26}).

For this purpose, we consider the Clauser-Horne-Shimony-Holt (CHSH) \cite{27}
Bell operator for
{\it polarization} ($\pi$) observables
\begin{equation}
\beta _{\pi }=-A_{\pi }\otimes B_{\pi }+A_{\pi }\otimes b_{\pi }+a_{\pi
}\otimes B_{\pi }+a_{\pi }\otimes b_{\pi }
\end{equation}
where the incompatible polarization observables of the first and
second photon are, respectively:
\begin{equation}
\begin{aligned}
A_{\pi }&=\left| H\right\rangle \left\langle H\right| -\left| V\right\rangle
\left\langle V\right|,\;\;
a_{\pi }=\left| V\right\rangle \left\langle H\right| +\left| H\right\rangle
\left\langle V\right|,\\
B_{\pi }&=\frac{1}{\sqrt{2}}\Bigl[\left| H\right\rangle \left\langle H\right|
-\left| V\right\rangle \left\langle V\right| +\left| V\right\rangle
\left\langle H\right| +\left| H\right\rangle \left\langle V\right| \Bigr]%
\text{,}\\
b_{\pi }&=\frac{1}{\sqrt{2}}\Bigl[\left| V\right\rangle \left\langle V\right|
-\left| H\right\rangle \left\langle H\right| +\left| V\right\rangle
\left\langle H\right| +\left| H\right\rangle \left\langle V\right| \Bigr]%
\text{,}
\end{aligned}
\end{equation}
They are chosen to provide a maximum violation (i.e., $2\sqrt{2}$)
of the CHSH-Bell inequality $\left| \beta _{\pi }\right| \leq 2$. In
addition, we consider the CHSH-Bell operator for {\it path} (${\bf
k}$) observables
\begin{equation}
\beta _{{\bf k}}=A_{{\bf k}}\otimes B_{{\bf k}}-A_{{\bf k}}\otimes b_{{\bf k}%
}+a_{{\bf k}}\otimes B_{{\bf k}}+a_{{\bf k}}\otimes b_{{\bf k}}
\end{equation}
where the (incompatible) path observables of the first and second
photon are, respectively:
\begin{equation}
\begin{aligned}
A_{{\bf k}}&=\left| l\right\rangle \left\langle r\right| +\left|
r\right\rangle \left\langle l\right| \quad,\quad
a_{{\bf k}}=i(\left| r\right\rangle \left\langle l\right| -\left|
l\right\rangle \left\langle r\right| )\text{,}\\
B_{{\bf k}}&=\left[ \left( i+1\right) \left| r\right\rangle \left\langle
l\right| -\left( i-1\right) \left| l\right\rangle \left\langle r\right| %
\right] /\sqrt{2}\;,\\
b_{{\bf k}}&=\left[ \left( i-1\right) \left| r\right\rangle \left\langle
l\right| -\left( i+1\right) \left| l\right\rangle \left\langle r\right| %
\right] /\sqrt{2}\;,
\end{aligned}
\end{equation}
They are chosen to provide a maximum violation (i.e., $2\sqrt{2}$)
of the CHSH-Bell inequality $\left| \beta _{{\bf k}}\right| \leq 2$.

In the second part of the experiment, we demonstrate that the
violation of the Bell inequalities grows exponentially with the
number of internal degrees of freedom. On this purpose, we consider
the $\pi-\bf k$ Bell operator
\begin{equation}
\beta =\beta _{\pi }\otimes \beta _{{\bf k}}\text{,}
\end{equation}
which requires measuring $4$ alternative $4$-outcome local observables on
each photon: $A_{\pi }A_{{\bf k}}$, $A_{\pi }a_{{\bf k}}$, $a_{\pi }A_{{\bf k}}$,
and $a_{\pi }a_{{\bf k}}$ on the first photon, and $B_{\pi }B_{{\bf k}}$,
$B_{\pi }b_{{\bf k}}$, $b_{\pi }B_{{\bf k}}$, and $b_{\pi }b_{{\bf k}}$ on
the second photon (i.e., we have $16$ different experimental
configurations). Each of the local observables is the product of a
polarization observable and a path observable. Any
theory admitting EPR elements of reality must satisfy the following inequality
\begin{equation}
\left| \beta \right| \leq 4\text{,} \label{beta}
\end{equation}
while quantum mechanics predicts a value of $8$ (with ideal
equipment). As a consequence, quantum nonlocality is expected to
grow exponentially with the number $N$ of degrees of freedom, i.e.
in our case the ratio between the quantum prediction and the
classical bound grows as $2^{\frac{N}{2}}$. The Bell inequality
(\ref{hyper-ent}) is based on Aravind's observation \cite{17} that
Mermin's growing-with-size quantum nonlocality effect also exists
for $n/2$ two-level Bell states. Under some assumptions \cite{22},
by hyper-entanglement we can replace n separated qubits by two
N-level systems.

Before observing the quantum violation of the two-degree of freedom Bell
inequality (\ref{beta}), we should test whether the assumptions leading to (\ref
{beta}) are satisfied in our experiment or not. These assumptions are:

($i$) The results of the measurements of each of the polarization and path
observables on photon $u$ ($d$) can be predicted with near certainty (i.e.,
with a sufficiently large probability) from the results of remote
measurements on photon $d$ ($u$) \cite{28}.

($ii$) If the same element of reality (for instance $A_{\pi }$) appears in
two different setups (for instance, $A_{\pi }A_{{\bf k}}$ and $A_{\pi }a_{{\bf k}}$),
the remote prediction for $A_{\pi }$ must be the same in both
setups \cite{28}.

To sum up, we must check that $A_{\pi }$ can be predicted (with
almost perfect certainty) from different experiments on $B$, and
that this prediction must not depend on whether $A_{\pi }$ is
measured using, for instance, the setup $A_{\pi }A_{{\bf k}}$ or the
setup $A_{\pi }a_{{\bf k}}$. This test requires measuring all
possible combinations of product local observables sharing one
polarization (or path) observable. For instance, the prediction for
the value $A_{\pi }$ on the first photon must be the same,
regardless of whether we measure $A_{\pi }A_{{\bf k}}$, or $A_{\pi
}a_{{\bf k}}$, or $A_{\pi }B_{{\bf k}}$, or $A_{\pi }b_{{\bf k}}$ on
the second photon, and regardless of whether we chose the setup
$A_{\pi }A_{{\bf k}}$, or $A_{\pi }a_{{\bf k}}$, or $A_{\pi }B_{{\bf
k}}$, or $A_{\pi }b_{{\bf k}}$ to measure $A_{\pi }$ on the first
photon (and so on). If ($i$) and ($ii$) are satisfied, then we can
proceed with the second part of the experiment and look for
violations of the two-degree of freedom Bell inequality
(\ref{beta}), in agreement with the quantum prediction.

The whole experiment admit two different analysis.
If we relax the EPR criterion and define
elements of reality as those that can be predicted with {\em almost}
perfect certainty, as in \cite{28}, then
the inequality (\ref{beta}) is valid for all prepared pairs, and
the experimental value of $\beta$ can be compared with the classical bound.
However, if we use the {\em original} EPR criterion, then the inequality (\ref{beta}) is legitimate
only for a fraction of pairs. Then, we should modify the classical bound of the inequality
in order to take into account the effect of the fraction of the pairs for which
the inequality is not valid \cite{22}.
In both approaches, a good measure of nonlocality is
the ratio between the experimental value of $\beta$ and
the maximal possible value allowed by the local
realistic theories. This measure is related both to the number of bits needed
to communicate nonlocally in order to emulate the experimental
results by a local realistic theory, and also to the minimum
detection efficiency needed for a loophole-free experiment. In this
sense, a higher value of this ratio is a
significant step towards a loophole-free Bell
test \cite{22}.

The source of entangled photons used in the experiment consists of a
thin type I $\beta$-barium-borate (BBO) crystal slab operating under
the double (back and forth) excitation of a cw $Ar^{+}$ laser
($\lambda _{p}=364$ nm)\cite{29}. The parametric source, which has
been described in detail in previous papers \cite{29,25}, is
schematically shown in Fig. 1a. Here we remember that the
polarization entangled states $|\Phi \rangle
=\frac{1}{\sqrt{2}}\left( |H\rangle _{u}|H\rangle _{d}+e^{i\theta
}|V\rangle _{u}|V\rangle _{d}\right) $ are obtained by superposition
of the two overlapping radiation cones of the crystal corresponding
to the degenerate wavelength, $\lambda =728$ nm. The phase $\theta $
of the state is controlled by micrometric displacements of the
mirror $M$ (see caption of Fig. 1).

\begin{figure}
\centerline{\includegraphics[width=9.7cm]{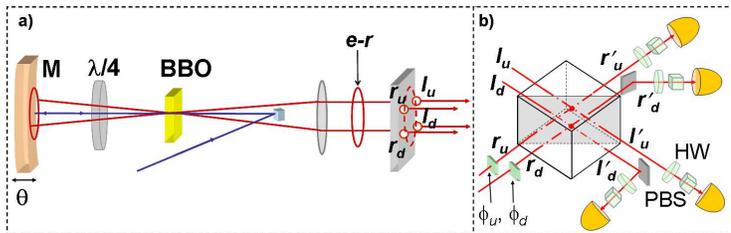}} \caption{a)
Parametric source of polarization-momentum hyper-entangled
two-photon states. The zero-order $\lambda /4$ waveplate placed
between $M\ $and the BBO\ crystal, intercepts twice both
back-reflected $\lambda $ and $\lambda _{p}$ beams and rotates by
$\pi /2$ the polarization of the back-reflected field at $\lambda \
$while leaving the polarization state of the UV\ pump beam virtually
undisturbed. The orthogonal section of the emission cone identifies
the {\it entanglement ring} ($e-ring$). Phase setting $\theta
=0,\pi$ are obtained by micrometric translation of $M$; b) Spatial
coupling of input mode sets $l_{u}-l_{d}$ and $r_{u}-r_{d}$ on the
$BS$ plane. The $BS$ output modes, $l_{u}^{\prime }$, $l_{d}^{\prime
}$, $r_{u}^{\prime }$, and $r_{d}^{\prime }$ are analyzed each by a
half-wave plate ($HW$) and a polarization beamsplitter ($PBS$).
Glass plates on modes $r_{u}$ and $r_{d}$ are used in order to
measure momentum observables.}
\end{figure}

Momentum entanglement is realized for either one of the two radiation cones
by selecting two pairs of correlated ${\bf k}$-modes, $l_{u}-r_{d}$ and $%
r_{u}-l_{d}$ within the $e-ring$ (Fig. 1a). Because of the
``phase-preserving'' character of the parametric process, the
relative phase between the two pair emissions is set to the value
$\phi =0$. Hence, for
each cone, the ${\bf k}$-entangled state $|\psi \rangle =\frac{1}{\sqrt{2}}%
\left( |l\rangle _{u}|r\rangle _{d}+|r\rangle _{u}|l\rangle
_{d}\right) $ can be generated. In the experiment, Bob and Alice's
sites are chosen in order to perform the measurements by the upper
($u$) and lower ($d$) detectors, respectively (cfr. Fig. 1b). The
mode sets $l_{u}-l_{d}$ and $r_{u}-r_{d}$ are spatially matched in
two different points of a symmetric beamsplitter $(BS)$: Fig. 1b. A
trombone mirror assembly mounted on a motorized translation stage
(not shown in the Figure) allows fine adjustments of the path delay
$\Delta x$\ between the input modes $l_{u}-l_{d}$ and $r_{u}-r_{d}$.
The photons associated with the output $BS$ modes, $l_{u}^{\prime
}-l_{d}^{\prime }$ and $r_{u}^{\prime }-r_{d}^{\prime }$, are
detected by four single photon detectors (Fig. 1b) within a
bandwidth $\Delta \lambda =6$ nm which corresponds to a
coherence-time of the down converted photons:\ $\tau _{coh}${\it \
}$\approx 150fsec $. Polarization analysis could be performed by
using a half-wave plate and a polarizing beamsplitter in each
detection arm. By varying the path delay around $\Delta x=0$, we
could observe a dip in the two-photon coincidences for the mode
combinations $l_{u}^{\prime }-r_{d}^{\prime }$ and $r_{u}^{\prime
}-l_{d}^{\prime }$, while a peak was observed in both cases:$\
l_{u}^{\prime }-l_{d}^{\prime }$, $r_{u}^{\prime }-r_{d}^{\prime }$
\cite{25}. The measured resonance "visibility" ($v\approx0 .90$) was
high enough to obtain a violation of the Bell inequality
(\ref{beta}).


\begin{table}[t]
{\begin{tabular}{c|cccc}
\hline \hline
& $A_{{\bf k}} B_{{\bf k}}$ & $A_{{\bf k}} b_{{\bf k}}$ & $a_{{\bf k}} B_{{\bf k}}$ & $a_{{\bf k}} b_{{\bf k}}$\\
\hline
$\mathbf{A}_{{\bf \pi }} \mathbf{A}_{{\bf \pi }}$ & $0.9077$ & $0.9097$ & $0.9054$ & $0.9133$
\\
$\mathbf{a}_{{\bf \pi }} \mathbf{a}_{{\bf \pi }}$ & $-0.8994$ & $-0.9128$ & $-0.8962$ & $-0.9069$ \\
$\mathbf{B}_{{\bf \pi }} \mathbf{b}_{{\bf \pi }}$ & $-0.8955$ & $-0.9033$ & $-0.9052$ & $-0.9008$ \\
$\mathbf{b}_{{\bf \pi }} \mathbf{B}_{{\bf \pi }}$ & $-0.9143$ & $-0.9206$ & $-0.9181$ & $-0.9196$ \\
\hline \hline
\end{tabular}}
\\ \vspace{0.2cm}
{\begin{tabular}{c|cccc}
\hline \hline
&$\mathbf{A}_{{\bf k}} \mathbf{A}_{{\bf k}}$ & $\mathbf{a}_{{\bf k}} \mathbf{a}_{{\bf k}}$ &
$\mathbf{B}_{{\bf k}} \mathbf{B}_{{\bf k}}$ & $\mathbf{b}_{{\bf k}} \mathbf{b}_{{\bf k}}$ \\
\hline
$A_{{\bf \pi }} B_{{\bf \pi }}$ & $0.8920$ & $0.8330$ & $0.8355$ & $0.8682$ \\
$A_{{\bf \pi }} b_{{\bf \pi }}$ & $0.8954$ & $0.8312$ & $0.8289$ & $0.8550$ \\
$a_{{\bf \pi }} B_{{\bf \pi }}$ & $0.8970$ & $0.8313$ & $0.8304$ & $0.8632$ \\
$a_{{\bf \pi }} b_{{\bf \pi }}$ & $0.8837$ & $0.8267$ & $0.8330$ & $0.8590$ \\
\hline \hline
\end{tabular}}
\caption{\small Above: experimental values of the polarization
observables $A_{{\bf \pi }}A_{{\bf \pi }}$, $a_{{\bf \pi }}a_{{\bf
\pi }}$, $B_{{\bf \pi }}b_{{\bf \pi }}$, and $b_{{\bf \pi }}B_{{\bf
\pi }}$ (bold) measured for different settings of the momentum
observables $A_{{\bf k}}B_{{\bf k}}$, $A_{{\bf k}}b_{{\bf k}}$,
    $a_{{\bf k}}B_{{\bf k}}$, and $a_{{\bf k}}b_{{\bf k}}$
(italics); below: experimental values of the momentum observables
$A_{{\bf k}}A_{{\bf k}}$, $a_{{\bf k}}a_{{\bf k}}$, $B_{{\bf
k}}B_{{\bf k}}$, and $b_{{\bf k}}b_{{\bf k}}$\ (bold) measured for
different settings of the polarization observables $A_{{\bf \pi
}}B_{{\bf \pi }}$, $A_{{\bf \pi }}b_{{\bf \pi }}$, $a_{{\bf \pi
}}B_{{\bf \pi }}$, and $a_{{\bf \pi }}b_{{\bf \pi }}$ (italics).
Experimental uncertainties are typically of the order of $0.0020$.}
\end{table}


Let us describe the different measurements performed in our
experiment. Every experimental result we show corresponds to a
measurement lasting an average time of $10\sec$. As a preliminary
step, we verified the violation of Bell inequalities on each degree
of freedom by separate measurements performed on polarization and
linear momentum by using the state (\ref{hyper-ent}). The
corresponding experimental results are: $\left| \left\langle \beta
_{\pi }\right\rangle \right| =2.5762\pm 0.0068$ and $\left|
\left\langle \beta _{{\bf k}}\right\rangle \right| =2.5658\pm
0.0067$ with a violation of local realism by $85$ and $84$ standard
deviations, respectively.

Before demonstrating the exponential growth of the Bell inequality
violation with $N=2$ of degrees of freedom of a $n=2$-photon
entangled state, we verified the preliminary theoretical
assumptions ($i$)
and ($ii$). Specifically, we performed the measurement of the operators $A_{%
{\bf \pi }}A_{{\bf \pi }}$, $a_{{\bf \pi }}a_{{\bf \pi }}$, $B_{{\bf \pi }%
}b_{{\bf \pi }}$, and $b_{{\bf \pi }}B_{{\bf \pi }}$, where the
first (second) operator refers to the $u$ ($d$) side. In order to
verify ($ii$), each measurement has been performed for any setting
of momentum analysis involved in the experiment, namely $A_{{\bf
k}}B_{{\bf k}}$, $A_{{\bf k}}b_{{\bf k}}$, $a_{{\bf k}}B_{{\bf k}}$,
and $a_{{\bf k}}b_{{\bf k}}$. The experimental results are shown in
Table I (above). An analogous procedure was followed for the
momentum observables $A_{{\bf k}}A_{{\bf k}}$, $a_{{\bf k}}a_{{\bf
k}}$, $B_{{\bf k}}b_{{\bf k}}$, and $b_{{\bf k}} B_{{\bf k}}$ [Table
I (below)]. The results in both Tables support both assumptions with
a reasonably high degree of certainty. The violation of the Bell
inequality (\ref{beta}) was demonstrated by performing
$16$ different simultaneous measurements of the polarization observables $A_{{\bf \pi }%
}B_{{\bf \pi }}$, $A_{{\bf \pi }}b_{{\bf \pi }}$, $a_{{\bf \pi }}B_{{\bf \pi
}}$, and $a_{{\bf \pi }}b_{{\bf \pi }}$, and the momentum observables $A_{{\bf k}}B_{%
{\bf k}}$, $A_{{\bf k}}b_{{\bf k}}$, $a_{{\bf k}}B_{{\bf k}}$, and $a_{{\bf k}%
}b_{{\bf k}}$ on both photons. The probabilities of each outcome for
the $16$ $\pi-\vec k$ settings are summarized in Table II.


\begin{table}[tbp]
\begin{center}
\begin{tabular}{c|cccc}
\hline \hline
& $\mathbf{A_{\bf k} B_{{\bf k}}}$ & $\mathbf{A_{{\bf k}} b_{{\bf k}}}$ &
$\mathbf{a_{{\bf k}} B_{{\bf k}}}$ & $\mathbf{a_{{\bf k}} b_{{\bf k}}}$ \\
\hline
$\mathbf{A}_{{\bf \pi }} \mathbf{B}_{{\bf \pi }}$ & $0.4210$ & $-0.4826$ &
$0.5080$ & $0.5108$ \\
$\mathbf{a}_{{\bf \pi }} \mathbf{B}_{{\bf \pi }}$ & $-0.5010$ & $0.4490$ & $-0.3254$ &
$-0.3792$ \\
$\mathbf{A}_{{\bf \pi }} \mathbf{b}_{{\bf \pi }}$ & $-0.4801$ & $0.3985$ & $-0.3205$ &
$-0.4431$ \\
$\mathbf{a}_{{\bf \pi }} \mathbf{b}_{{\bf \pi }}$ & $-0.4508$ & $0.4536$ & $-0.4479$ &
$-0.4475$ \\
\hline \hline
\end{tabular}
\end{center}
\caption{\small Experimental values of the $16$ different joint
measurements of the polarization, $A_{{\bf \pi }}B_{{\bf \pi }}$, $A_{{\bf %
\pi }}b_{{\bf \pi }}$, $a_{{\bf \pi }}B_{{\bf \pi }}$, and $a_{{\bf \pi }}b_{%
{\bf \pi }}$, and the momentum observables $A_{{\bf k}}B_{{\bf k}}$, $A_{{\bf k}%
}b_{{\bf k}}$, $a_{{\bf k}}B_{{\bf k}}$, and $a_{{\bf k}}b_{{\bf
k}}$ performed on both photons. Experimental uncertainties are
typically of the order of $0.0040$. The exact values will be
reported shortly in a more extended paper.}
\end{table}


The experimental value of $\left| \beta \right|$, obtained after summation
over all the measured values of Table II, $\left| \left\langle \beta
\right\rangle \right| =7.019\pm0.015$, corresponds to a violation of the
inequality (\ref{beta}) by $196$ standard deviations, demonstrating the
magnitude of the contradiction with local realism achievable with the
$2$-photon hyper-entangled state (\ref{hyper-ent}).
Assuming the version of EPR elements of reality proposed
in \cite{28}, we have obtained an experimental value of $\left| \left\langle \beta
\right\rangle \right|/4 = 1.7548$ for two degrees of freedom (polarization
and path) vs an experimental value of $\left| \left\langle \beta
_{\pi }\right\rangle \right| /2 = 1.2881$ for polarization, and
$\left| \left\langle \beta _{{\bf k}}\right\rangle \right| /2=
1.2829$ for path.


In this Letter we have given the first experimental demonstration of
Mermin's prediction that the nonlocal character of a quantum state grows
with the dimension of the Hilbert space \cite{9}. The experiment has been
performed by using a polarization momentum two-photon hyper-entangled state.
A further extension to a larger Hilbert space could show an even more
significant deviation from classical bounds by entangling both particles
in other degrees of freedom of the hyper-entangled state. We are presently
investigating in our Laboratory the adoption of time bin for this purpose.

This work was supported by the FIRB 2001
and PRIN 2005 
of MIUR (Italy) and by the FET European Network on Quantum
Information and Communication, and the Spanish MEC Project No.
FIS2005-07689.



\begin{references}
\bibitem{1} A. Einstein, B. Podolsky, and N. Rosen, Phys. Rev. {\bf 47}, 777 (1935).

\bibitem{2} J. S. Bell, Physics (Long Island City, NY) 1, 195 (1964).

\bibitem{3} N. D. Mermin, Phys. Rev. D {\bf 22}, 356 (1980).

\bibitem{4} A. Garg and N. D. Mermin, Phys. Rev. Lett. {\bf 49}, 901
(1982); {\bf 49}, 1294 (1982).

\bibitem{5} A. Garg and N. D. Mermin, Phys. Rev. D {\bf 27}, 339
(1983).

\bibitem{6} N. Gisin and A. Peres, Phys. Lett. A {\bf 162}, 15 (1992).

\bibitem{7} D. G. Collins, N. Gisin, N. Linden, S. Massar, S. Popescu,
Phys. Rev. Lett. {\bf 88}, 040404 (2002).

\bibitem{8} J.-L. Chen, D. Kaszlikowski, L. C. Kwek, C. H. Oh, and M. \.{Z}ukowski,
Phys. Rev. A {\bf 64}, 052109 (2001).

\bibitem{9} N. D. Mermin, Phys. Rev. Lett. {\bf 65}, 1838 (1990).

\bibitem{10} D. M. Greenberger, M. A. Horne, and A. Zeilinger, in {\it
Bell's Theorem, Quantum Theory, and Conceptions of the Universe},
edited by M. Kafatos (Kluwer, Dordrecht, 1989), p. 69.

\bibitem{11} D. Bouwmeester, J.-W. Pan, M. Daniell, H. Weinfurter, and A.
Zeilinger, Phys. Rev. Lett. {\bf 82}, 1345 (1999).

\bibitem{12} J.-W. Pan, D. Bouwmeester, M. Daniell, H. Weinfurter, and A.
Zeilinger, Nature (London) {\bf 403}, 515 (2000).

\bibitem{13} J.-W. Pan, M. Daniell, S. Gasparoni, G. Weihs, and A. Zeilinger,
Phys. Rev. Lett. {\bf 86}, 4435 (2001).

\bibitem{14} Z. Zhao, T. Yang, Y.-A. Chen, A.-N. Zhang, M. \.{Z}ukowski, J.-W.
Pan, Phys. Rev. Lett. {\bf 91}, 180401 (2003).

\bibitem{15} M. Hein, W. D\"{A}ur, and H.-J. Briegel, Phys. Rev. A
{\bf 71}, 032350 (2005).

\bibitem{16} A. Cabello, Phys. Rev. Lett. {\bf 87}, 010403 (2001).

\bibitem{17} P. K. Aravind, Found. Phys. Lett. {\bf 15}, 397 (2002).

\bibitem{18} A. Cabello, Phys. Rev. Lett. {\bf 95}, 210401 (2005).

\bibitem{19} A. Cabello, Phys. Rev. A {\bf 72}, 050101(R) (2005).

\bibitem{20} C. Cinelli, M. Barbieri, R. Perris, P. Mataloni, and F. De
Martini, Phys. Rev. Lett. {\bf 95}, 240405 (2005).

\bibitem{21} T. Yang, Q. Zhang, J. Zhang, J. Yin, Z. Zhao, M. \.{Z}ukowski,
Z.-B. Chen, and J.-W. Pan, Phys. Rev. Lett. {\bf 95}, 240406 (2005).

\bibitem{22} A. Cabello, ``Bell Inequalities Based on Equalities'' (work in progress).

\bibitem{23} P. G. Kwiat, J. Mod. Opt. {\bf 44}, 2173 (1997).

\bibitem{24} P. G. Kwiat and H. Weinfurter, Phys. Rev. A {\bf 58},
R2623 (1998).

\bibitem{25} M. Barbieri, C. Cinelli, F. De Martini, and P. Mataloni,
Phys. Rev. A {\bf 72}, 052110 (2005).

\bibitem{26} J. T. Barreiro, N. K. Langford, N. A. Peters, and P. G. Kwiat,
Phys. Rev. Lett. {\bf 95}, 260501 (2005).

\bibitem{27} J. F. Clauser, M. A. Horne, A. Shimony, and R. A. Holt, Phys.
Rev. Lett. {\bf 23}, 880 (1969).

\bibitem{28} P. H. Eberhard and P. Rosselet, Found. Phys. {\bf 25},
91 (1995).

\bibitem{29} C. Cinelli, G. Di Nepi, F. De Martini, M. Barbieri, and P.
Mataloni, Phys. Rev. A {\bf 70}, 022321 (2004).

\end{references}
\end{document}